\documentclass[final] {aipproc}
\usepackage{sidecap,graphicx,boxedminipage,epsfig,wrapfig,floatflt,shadow}

\layoutstyle{8x11double}

\begin{document}

\title{ 
Assessing Student Expertise in Introductory Physics with Isomorphic
Problems, Part II: Effect of Some Potential Factors on Problem Solving and Transfer
}


\author{Chandralekha Singh}{
  address={Department of Physics and Astronomy, University of Pittsburgh, Pittsburgh, PA, 15260}}

\begin{abstract}
In this companion paper, we explore the use of isomorphic problem
pairs (IPPs) to assess introductory physics students' ability to solve
and successfully transfer problem-solving knowledge from one context to another in
mechanics. We call the paired problems {}``isomorphic'' because
they require the same physics principle to solve them. 
We analyze written responses and individual discussions to a range of 
isomorphic problems. We examine potential factors that may help or
hinder transfer of problem-solving skills from one problem in a pair
to the other. For some paired isomorphic problems, one context often
turned out to be easier for students in that it was more often correctly
solved than the other. When quantitative and conceptual questions
were paired and given back to back, students who answered both questions
in the IPP often performed better on the conceptual questions than
those who answered the corresponding conceptual questions only. Although
students often took advantage of the quantitative counterpart to answer
a conceptual question of an IPP correctly, when only given the conceptual
question, students seldom tried to convert it into a quantitative
question, solve it and then reason about the solution conceptually.
Even in individual interviews when students who were only given conceptual
questions had difficulty and the interviewer explicitly encouraged
them to convert the conceptual question into the corresponding quantitative
problem by choosing appropriate variables, a majority of students
were reluctant and preferred to guess the answer to the conceptual
question based upon their gut feeling. 
Misconceptions associated with friction in some problems were so robust that pairing them with isomorphic
problems not involving friction did not help students discern their underlying similarities.
Alternatively, from the knowledge in pieces perspective, the activation of the knowledge resource
related to friction was so strongly and automatically triggered by the context, which is outside the conscious control
of the student, that students did not look for analogies with paired problems or other aids that may be present.
\end{abstract}
 
\maketitle

\section{Introduction}


In this companion paper, we explore the use of isomorphic problem
pairs (IPPs) to assess introductory physics students' expertise in
mechanics in a range of contexts. We call the paired problems isomorphic
if they require the same physics principle to solve them. We investigate
a few parameters as potential factors that may help problem solving
and analyze the performance of students on the IPPs from the perspective
of {}``transfer\char`\"{}~\cite{gick,mestre,lobato,sanjay,bassok,bransford,sternberg,transfer5,analogy_transfer,hendrickson,klahr}.
For example, we examine the effect of misconceptions about friction
as a potential barrier for problem solving and {}``transfer\char`\"{}.~\cite{gick,mestre,lobato,sanjay}
Transfer in physics is particularly challenging because there are
only a few principles and concepts that are condensed into a compact
mathematical form. Learning requires unpacking them and understanding
their applicability in a variety of contexts that share deep features,
e.g., the same law of physics may apply in different contexts. Cognitive
theory suggests that transfer can be difficult especially if the {}``source\char`\"{}
(from which transfer is intended) and the {}``target\char`\"{} (to
which transfer is intended) do not share surface features. This difficulty
arises because knowledge is encoded in memory with the context in
which it was acquired and solving the source problem does not automatically
manifest its {}``deep\char`\"{} similarity with the target problem.~\cite{gick}
Ability to transfer relevant knowledge from one context to another improves with expertise because an expert's knowledge
is hierarchically organized and represented at a more abstract level
in memory, which facilitates categorization and recognition based
upon deep features.~\cite{mestre,lobato,sanjay,bassok,bransford,sternberg,transfer5,analogy_transfer,chi3,hardiman}

Students may find one problem in an IPP easier to tackle than its
pair because context and representation are very important.~\cite{mestre,lobato,sanjay,bassok,bransford,sternberg,transfer5,analogy_transfer,chi3,hardiman}
For example, if two equivalent groups of students are given only the quantitative or conceptual
question from an IPP pairing a quantitative and conceptual question
with similar contexts, one group may perform well on one of them but not
the other. Some studies have shown that if students are reasonably
comfortable with mathematical manipulation required to solve a quantitative
problem, the group given the quantitative problem may perform better on it using an algorithmic approach
than the group given the corresponding conceptual question~\cite{lillian,kim,eric}.
In a study on student understanding of diffraction and interference
concepts, the group which was given a quantitative problem performed significantly better
than the group given a similar conceptual question~\cite{lillian}. In another study
Kim et al.~\cite{kim} examined the relation between traditional
physics textbook style quantitative problem solving and conceptual
reasoning. They found that, although students in a mechanics course
on average had solved more than 1000 quantitative problems and
were facile at mathematical manipulations, they still had many common
difficulties when answering conceptual questions on related topics.
When Mazur~\cite{eric} gave a group of Harvard students quantitative problems
related to power dissipation in a circuit, students performed significantly
better than when an equivalent group was given conceptual questions about the relative
brightness of light bulbs in similar circuits. In solving the quantitative
problems given by Mazur, students applied Kirchhoff's rules to write
down a set of equations and then solved the equations algebraically
for the relevant variables from which they calculated the power dissipated.
When the conceptual circuit question was given to students in similar
classes, many students appeared to guess the answer rather than reasoning
about it systematically~\cite{eric}. For example, if students are
given quantitative problems about the power dissipated in each headlight
of a car with resistance $R$ when both bulbs are connected in parallel
to a battery with an internal resistance $r$ and then asked to repeat
the calculation for the case when one of the headlights burned out,
the procedural knowledge of Kirchhoff's rules can help students solve
for the power dissipated in each headlight even if they cannot conceptually
reason about the current and voltage in different parts of the circuit~\cite{lillian,eric}.
To reason without resorting \textit{explicitly} to mathematical tools
(Kirchhoff's rules) that the single headlight in the car will be brighter
when the other headlight burned out, students will have to reason in the following
manner: The equivalent resistance of the circuit is lower when both headlights
are working so that the current coming out of the battery is larger.
Hence, more of the battery voltage drops across the internal resistance
$r$ and less of the battery voltage drops across each headlight
and therefore each headlight will be less bright. If a student deviates from this
long chain of reasoning required in conceptual reasoning, the student
may not make a correct inference.

\vspace*{-0.2in}
 
\section{Hypotheses and Goals}

\vspace*{-0.2in}

The experiments we describe here can broadly be classified into three categories.
Experiment 1 involves IPPs which pair a quantitative question with a conceptual
question. Experiment 2 involves IPPs in which both questions are conceptual.
Experiment 3 addresses the effect of misconceptions about friction on students'
ability to transfer relevant knowledge from a problem not involving friction
to isomorphic problems involving friction.

We developed several IPPs in the multiple-choice format (final version shown in the Appendix)
with different contexts in mechanics. The problems spanned a range of difficulty. 
The correct solution to each question 
is italicized in the Appendix.
We administered either one or both questions in an IPP to introductory
physics students. We made hypotheses H1-H3 related to experiment 1, hypothesis H4 related
to experiment 2, and hypothesis H5 related to experiment 3 as described below:

\begin{itemize}
\item Experiment 1 with IPPs in which one question is more quantitative
than the other: Although it is difficult to categorize physics questions
as exclusively quantitative or conceptual, some of the IPPs had one
question that required symbolic or numerical calculation while the
other question could be answered by conceptual reasoning alone. The
first five IPPs in the Appendix fall in this category (although questions
(3) and (4) in the second IPP can both be classified as quantitative). We
made the following hypotheses regarding these IPPs:

\begin{itemize}
\item H1: 
Performance on quantitative questions of an IPP will be better when both the quantitative and conceptual questions
are given than when only the quantitative question is given.
\item H2: 
Performance on conceptual questions of an IPP will be better when both the quantitative and conceptual questions
are given than when only the conceptual question is given.
\item H3: The closer the match between the contexts
of the quantitative and conceptual questions of an IPP, the better will students 
be able to discern their similarity and transfer relevant knowledge from one problem to another.
\end{itemize}

We note that our study is different from those mentioned earlier~\cite{lillian,kim,eric} because
our goal is not to evaluate whether students perform better on the quantitative or
conceptual question but rather to evaluate whether giving
both questions together improves performance on each type of question compared to the case when only the conceptual or
the quantitative question alone was given.

Hypothesis H1 is based on the assumption that solving the conceptual question of an IPP may 
encourage students to perform a qualitative analysis, streamline students' thinking, 
make it easier for them to narrow down relevant concepts, and thus help them solve the quantitative problem correctly.
Prior studies show that introductory physics students are not systematic in using effective problem solving strategies, and often do not perform
a conceptual analysis while solving a quantitative problem~\cite{tutorial}. 
They often use a ``plug and chug" approach to solving quantitative problems which may prevent them from solving the problem correctly.
The conceptual questions may provide an opportunity for reflecting upon the quantitative problem and performing a qualitative analysis and planning.
This can increase the probability of solving the quantitative problem correctly.
We note that since the IPPs always had a quantitative question preceding the corresponding conceptual question, hypothesis
H1 assumes that students will go back to the quantitative question if they got some insight from the corresponding conceptual
question.

Hypothesis H2 is inspired by results of prior studies that show that introductory physics students often perform 
better on quantitative problems compared to conceptual questions on the same topic~\cite{lillian,kim,eric}.
Students often treat conceptual questions as guessing tasks~\cite{lillian,kim,eric}.
We hypothesized that students who are able to solve the quantitative problem in an IPP may use its solution as a hint for answering
the conceptual question correctly if they are able to discern the similarity between the two questions.
Since quantitative and conceptual questions of an IPP were given one after another,
we hypothesized that students would likely discern their underlying similarity at least in cases where the contexts were similar. 
When reasoning without quantitative tools, it may be more difficult to create the correct chain of reasoning
if a student is ``rusty" about a concept.~\cite{fred2} Equations can provide a pivot point for constructing the reasoning chain.
For example, if a student has forgotten whether the maximum safe driving speed while making a turn
on a curved road depends on the mass of the vehicle, he/she will have great difficulty
reasoning without equations that the maximum speed is not dependent on the mass.
Similarly, a student with evolving expertise who is comfortable reasoning with equations may need to
write down Newton's second law {\it explicitly}
to conclude that the tension in the cable of an elevator accelerating upward is greater than its weight.
An expert can use the same law {\it implicitly} and conceptually argue that the upward acceleration implies that
the tension exceeds the weight without writing down Newton's second law explicitly.
Being able to reason conceptually without resorting to quantitative tools in a wide variety of contexts
may be a sign of adaptive expertise whereas conceptual reasoning by resorting to quantitative tools
may be a sign of evolving expertise~\cite{fred2,hatano}.

Hypothesis H3 is based upon results of prior studies related to
transfer.~\cite{gick,mestre,lobato,sanjay,bassok,bransford,sternberg,transfer5,analogy_transfer,hendrickson,klahr} For example, in
teaching debugging in Logo programming to children and investigating near and far transfer of debugging skills to other contexts, 
Carver et al. found that transfer of relevant knowledge is easier if the contexts of the problems are similar~\cite{klahr}. 
In the IPPs with paired questions with different contexts, transfer of relevant knowledge may be
more difficult because students may have more difficulty discerning
their underlying similarity. If the contexts are very different, discerning
the underlying similarity of the problems in each pair can be considered a sign of adaptive expertise~\cite{hatano}. 
If students had difficulty discerning the underlying similarity of the IPPs with different contexts, we explore the aspects
of the IPPs that made the transfer of relevant knowledge difficult.
Amongst the first five IPPs, we identified the contexts of the questions in the first three IPPs to be closest, followed by the IPP pairing
questions (9)-(10), and then the IPP pairing questions (7)-(8).
The main difference between the contexts of questions in IPP (9)-(10) is that in one case a person is falling vertically into a boat moving 
horizontally and in the other case rain is falling vertically into a cart moving horizontally. 
The IPP with questions (7) and (8) was considered to be the one requiring farthest transfer of relevant knowledge because the quantitative problem
(7) asks about the time for a projectile to reach the maximum height and question (8) asks students to compare the time of flight for three
projectiles launched with the same speed that achieved different heights and had different horizontal ranges. In order to transfer from
problem (7) to (8), students need to know that the total time of flight for a projectile is twice the time to reach the maximum height.
Moreover, in question (8), students should not get distracted by different horizontal ranges for the three projectiles, since the horizontal
range is not a relevant variable for answering this question.

\item
Experiment 2 involves IPPs with different contexts in which 
neither question is quantitative: Examples of three such IPPs are pairs in questions (11)-(16) in the Appendix.
We made the following hypothesis:

\begin{itemize}
\item H4: When both questions of an IPP are conceptual, performance will be better when both questions are given versus when only one is given.
\end{itemize}

Hypothesis H4 is based upon the assumption that one question in an IPP may provide a hint for the other question and may
help students in converging their reasoning based upon relevant principles and concepts.

\item Experiment 3 involves IPPs or a problem triplet in which some questions involve distracting features, e.g., 
common misconceptions related to friction. 
IPPs involving questions (18) and (20), questions (24) and (25), and the triplet involving questions (21), (22) and (23)
in the Appendix address this issue. We made the following hypothesis:

\begin{itemize}
\item H5: In IPPs or problem triplets where some questions are related to friction for which misconceptions are prevalent, 
performance will be worse on the friction question than on the question that does not contain friction. Giving such problems
involving friction with isomorphic problems not involving friction will not improve performance on the problems with friction.
\end{itemize}

Hypothesis H5 is based upon the assumption that distracting features such as misconceptions can divert students'
attention away from the central issue and may mask the similarity between questions in an IPP. From the perspective of knowledge in pieces,
problem context with distracting features can trigger the activation of knowledge that a student thinks is relevant but which is not actually 
applicable in that context.
The student may feel satisfied applying the activated knowledge resource and may not look further for analogies to paired problems or other aids.
Thus, transfer of relevant knowledge in these cases may be challenging.
One common misconception about the static frictional force is that it is always at its maximum
value, because students have difficulty with the mathematical inequality
that relates the magnitude of the static frictional force with the
normal force.~\cite{disessa} Students overgeneralize the inequality
regarding the static frictional force and think that since we so often set static friction
to its maximum value, it must be maximum all the time. Students also have difficulty in determining
the direction of the frictional force. Another difficulty students have is in determining
when static vs. kinetic friction is relevant for a problem.  

\end{itemize}

\vspace*{-0.2in}

\section{Methodology}

\vspace*{-0.2in}

Students in nine college calculus-based introductory physics courses participated in the study.
The questions were asked after instruction in relevant concepts and after students
had an opportunity to work on their homework on related topics. Some
students were given both questions of an IPP (or all three questions 
in triplet questions (21)-(23)), which were asked back
to back, while others were given only one of the two questions. 
When students were given both questions of an IPP back to back, the questions
were always given in the order given in the Appendix. 
For example, in the first five IPPs, the quantitative questions preceded the corresponding
conceptual question. 
However, students were free to go back and forth between them if they wished and could change
the answer to the previous question if they acquired additional insight
for solving the previous question by answering a latter question.
Students who were given both questions of an IPP were \underline{not} told
explicitly that the questions given were isomorphic. Students were given
2.5 minutes on an average to answer each question.

Not all of the IPPs were used for all of the eight courses due to
logistical difficulties. In particular, instructors of the courses
often were concerned about the time it would take to administer all
of the questions and they ultimately determined which questions from
the IPPs they gave to their classes. In some cases, depending
upon the consent of the course instructor (due to the time constraint
for a class), students were asked to explain their reasoning in each
case to obtain full credit. The questions contributed to students'
grades in all courses. In some of the courses, we discussed the responses
individually with several student volunteers. In one of these courses,
students were given a survey after they had worked on the IPPs to
evaluate the extent to which they realized that the questions were
isomorphic and how often they took advantage of their response to
one of the questions to solve its pair. Because the patterns of student
responses are similar for different classes, we discuss the responses
collectively here.

\vspace*{-0.2in}
 
\section{Results and Discussion}

\vspace*{-0.2in}

For the isomorphic problems given in the multiple-choice format in
the Appendix, Table 1 summarizes the numbers of students who were given
both questions or one of the questions of an IPP (or all three questions (21)-(23)), and students' average performance. 
Table 1 also shows the results of a Chi-square test with both the $\chi^{2}$ and $p$ values for comparison between
cases when both questions in an IPP (or all three questions (21)-(23)) were given vs. only one of the questions was given.
Students can make appropriate connections between the questions in an IPP only if they
have a certain level of expertise that helps them discern the connection
between the isomorphic questions. Improved student performance when both
questions of an IPP were given vs. when only one of the questions
was given was taken as one measure of transfer of relevant knowledge from one problem to another.
Below we discuss the findings and analyze student performance in light of our hypotheses H1-H5.

\subsection{Experiment 1: IPPs with Quantitative/Conceptual Pairs}

Table 1 shows that contrary to our hypothesis H1, student performance on quantitative questions was not significantly different 
when both quantitative and conceptual questions were given back to back (with the quantitative question
preceding the conceptual question) than when only the corresponding quantitative question was given. 
In some cases, the performance on the conceptual question was better than the performance on the quantitative
question (problem pairs (9)-(10)), but students could not leverage their conceptual knowledge for gain on the corresponding
quantitative problem. As noted earlier, the two questions in an IPP were
always given in the same order although students could go back and forth
if they wanted. It is possible that students overall did not go back to
the questions they had already answered, especially due to the time constraint, even if the question that followed
provided a hint for it. Future research will evaluate the effect of
switching the order of the quantitative and conceptual questions in an IPP when both questions are given. 

On the other hand, in support of hypothesis H2, students who worked on both questions of the IPPs
involving a conceptual and a quantitative problem performed better
on the conceptual questions at least for three of the five IPPs than when they were given only the conceptual questions. 
Table 1 shows that, for three of the IPPs, if one question in an IPP was quantitative and another
conceptual (the first five problem pairs), students often performed better on the conceptual question
when both questions were given rather than the corresponding conceptual question alone.
The fact that many students took advantage of the quantitative problem to solve the conceptual question points to their evolving expertise. 
For example, many students who were given both questions (1) and (2) recognized that the final momenta of the ships 
are independent of their masses under the given conditions by solving the quantitative problem. 
Written responses and individual discussions suggest that some students who
answered the conceptual question (2) correctly were not completely
sure about whether the change in momentum in question (1) was given
by option (a) or (e). However, since the answer in either case is
independent of the mass of the object, these students chose the correct
option (c) for question (2). The students who chose the incorrect
option (a) for question (1) but the correct option (c) for question
(2) often assumed that both ships in question (2) must have traveled
the same distance although that is not correct. 
In individual discussions, several students explicitly noted that
the mass cancels out in question (5) so the answer to question (6) cannot depend on mass. 
Similarly, discussions with individual students and students' written work suggest that solving the
quantitative question (9) helped many students formulate their solution
to question (10). Although some students were not able to solve the
quantitative question, e.g., due to algebraic error or not realizing
that when considering the conservation of the horizontal component of
momentum, Batman's vertical velocity should not be included, it was
easier for them to answer the conceptual question after thinking about
the quantitative one. Most of them realized that the boat would slow down after Batman lands in it. 

Previous research shows that answering conceptual questions can sometimes be more challenging for students than quantitative
ones, if the quantitative problems can be solved algorithmically and
students' preparation is sufficient to perform the mathematical manipulations.~\cite{lillian,kim,eric,fred}
If a student knows which equations are involved in solving a quantitative
problem or how to find the equations,
he or she can combine them in \textit{any order} to solve
for the desired variables even without a deep conceptual understanding
of relevant concepts. On the contrary, while reasoning without equations,
the student must usually proceed in a \textit{particular order} in
the reasoning chain to arrive at the correct conclusion~\cite{lillian,kim,eric,fred}.
Therefore, the probability of deviating from the correct reasoning
chain increases rapidly as the chain becomes long. We note however
that our hypothesis H2 is not about whether students will perform better
on the quantitative or conceptual question of an IPP when the
two questions are given separately (especially because the wording is not parallel for the
quantitative and conceptual questions in an IPP). Rather, our hypothesis
relates to whether students will recognize the similarity of the quantitative
and conceptual questions in an IPP, and take advantage of their solution to one question to answer the corresponding paired question.
Our finding suggests that students can leverage their quantitative solutions to correctly answer the corresponding
conceptual questions.

The fact that students often performed better on conceptual questions
when they were paired with quantitative questions brings up the following
issue. If students could turn the conceptual questions into analogous
quantitative problems themselves when only the conceptual questions
were given, they may have solved the quantitative problem algorithmically
if they were comfortable with the level of mathematics needed, and
then reasoned qualitatively about their results to answer the original
conceptual question. Almost without exception, students did not do this.
One can hypothesize that students have not thought seriously about
the fact that a conceptual question can be turned into a quantitative
problem, or that a mathematical solution can provide a tool for reasoning
conceptually. Without explicit guidance, students may not realize
that this conversion route may be more productive than carrying out
long conceptual reasoning without mathematical relations. However,
we find that students avoided turning conceptual questions into quantitative
ones, even when explicitly encouraged to do so. In one-on-one interview
situations, when students were only given the conceptual questions,
they also tried to guess the answer based upon their gut feeling.
More research is required to understand why students are reluctant
to transform a conceptual question into a quantitative problem even
if the mathematical manipulations required after such a conversion
and making correct conceptual inferences are not too difficult for them. One
possible explanation for such reluctance is that such a transformation
from a conceptual to a quantitative problem is cognitively demanding for
a typical introductory physics student and may cause a mental overload.~\cite{sweller}
According to Simon's theory of bounded rationality, an individual's
rationality in a particular context is constrained by his/her expertise
and experience and an individual will only choose one of the few options
consistent with his/her expertise that does not cause a cognitive overload.~\cite{simon}

Consistent with hypothesis H3, student performance on question (8) 
did not improve significantly when it was given together with question (7).
Discussions with individual students who answered both questions (7) and (8) suggest
that after solving the quantitative problem, some students were unsure whether the horizontal component of
motion was important or not. Students also needed to know that the time to reach the maximum height is half of the time of flight.
However, students' performance on question (10) improved significantly when it was given with question (9) rather than alone despite
the fact that the contexts were somewhat different; in particular, in one case a person is falling vertically and in the other case rain is falling 
vertically. In this case, many students were able to transfer relevant knowledge from question (9) to (10).

We note that for the IPP in questions (3) and (4), the quantitative problem itself was very challenging.
Most interviewed students and those who wrote something on their answer sheet did not
use conservation of energy correctly and forgot to take into account
both the rotational and translational kinetic energies in their analysis.
Thus, it is not surprising that there is no significant difference between cases when only one
of the questions was given vs. both questions were given.

Some of the quantitative questions asked for numerical answers while others asked for symbolic answers.
Individual discussions suggest that students were often able to take advantage of their process for quantitative
solution in either case to tackle the conceptual question more successfully (e.g., question (1) asked for a symbolic answer
whereas question (5) asked for a numerical answer) than if they were only given the conceptual question.
Future research will further investigate the differences in numerical vs. symbolic answers by giving
identical questions requiring numerical answers from some students and symbolic answers from others.

\subsection{Experiment 2: IPPs that do not mix quantitative/conceptual questions}

Table 1 shows that, in support of hypothesis H4, students' performance often improved when both questions of an IPP
were given compared to when only one of the two questions was given.
For example, Table 1 shows that the performance on both questions (11) and (12) improved when both questions were given.
Individual interviews and written responses suggest that students sometimes got confused about the distinction
between angular momentum and angular speed. However, students who answered question (11) correctly were often
able to extend their argument to question (12) and they were able to identify that the angular momentum does not change
and angular speed increases when the star collapses.
For example, during an interview, a student who answered both questions (11) and (12), first narrowed
down the possible correct choices for question (12) to (a) or (e),
noting that the angular momentum does not change here, similar to the skater
problem. Then the student noted that, since the angular speed must
increase as the star shrinks, the correct choice must be (e). This
student took clues from the question about the ice skater and answered
the white dwarf question correctly, explicitly making the comparison
between the paired questions and quickly eliminating options (b),
(c) and (d) in question (12), which sheds light on this student's expertise
and his ability to transfer relevant knowledge from one context to another.

Similarly, in question (14), in which ball B is unconventional
in that the density is not uniform and the inner core is denser than the outer shell,
student performance improved when it was given with question (13).
Individual discussions and the difference between the correct responses for the cases where students answered both questions
on the IPP vs. only question (14) suggest that students took advantage
of the scaffolding provided by the paired problem. In particular,
question (13) specifically helped students to consider whether the
mass and the radius are the relevant variables, or the moment of inertia
(the distribution of mass). Many students appeared to have the expertise
to transfer this knowledge to question (14). On the other hand, question (14) does not provide any hints for question (13)
and appears not to be helpful for answering question (13) when both questions were given as a pair.

In the IPP involving questions (15) and (16), students performed significantly better on question (16) when both questions of the IPP was given. 
It is somewhat surprising that students did better on the turntable problem than the bandit problem when both were given.
Some students who were given both questions (15) and (16) claimed that the speed of the cart will be unchanged after the bandit lands in it.
It may be due to the fact that students did not recognize the bandit question as a completely
inelastic collision problem, for which the object slows down after the
collision. This is not surprising considering the \textit{subtle} fact that, in the bandit problem,
there must be a frictional force between the bandit and the cart which
will slow down the cart and bring the bandit to the same horizontal speed
as the cart. In particular, after the bandit falls in the cart, the frictional force
that the cart exerts on bandit's shoes is equal in magnitude but opposite
in direction to the force that the bandit's shoes exert on the cart from Newton's third law. These forces will slow
the cart down and speed up the bandit so that they both have the same final horizontal velocity. 
In a typical inelastic collision problem given to students,
objects do not move perpendicular (but parallel) to each other before
the collision, as in the case of two cars colliding head-on and sticking to each other. 
In the bandit problem, some interviewed students and those providing written explanation reasoned 
incorrectly that the vertical motion of the bandit cannot affect the horizontal motion of the cart.
This misconception originates from the decoupling of the vertical and horizontal motion, e.g., for a projectile. 
Interviews suggest that for the turntable problem students used their intuition and experience about
this problem to predict that the turntable will slow down when the
putty falls on it and did not explicitly invoke conservation of angular
momentum. Future research will involve giving these problems in the
opposite order to evaluate the ordering effect.

\subsection{Experiment 3: Influence of distracting features and misconceptions about friction}

Consistent with hypothesis H5, students had difficulty in seeing the deep connection between
the isomorphic problems not involving friction and those involving friction~\cite{disessa},
even though they were given back to back, and in transferring relevant knowledge to the problem involving friction. 
The fact that students did not take advantage of the easier problems not involving friction to answer
the questions involving friction suggests that the misconceptions
about friction were quite robust.~\cite{friction} Many students believed that (i) the
static friction is always at the maximum value, (ii) the kinetic friction
is responsible for keeping the car at rest on an incline, or (iii) the presence or absence
of friction must affect the work done by {\bf you} even if you apply the same force over the same distance. 

In the IPP involving questions (18) and (20), the weight of the car and the normal force exerted on the car by the inclined
surface are the same in both problems. The only other force acting
on the car (which is the tension force in one problem and the static frictional force in the other problem) must be the same. 
Consistent with the common misconception about the static frictional force that it must be at its maximum value $f_{s}^{max}=\mu_{s}N$, 
(where $\mu_{s}$ is the coefficient of static friction and $N$ is the magnitude of the normal force), 
the most common incorrect response to question (20) was $\mu_{s}N=11,700$ N ($\sim40\%$).
Giving both questions (18) and (20) did not improve student performance on question (20) compared to when it was given alone. 

In order to help students discern the similarity between questions
(18) and (20), we later introduced two additional questions (17) and (19) that asked
students to identify the correct free body diagrams for questions (18) and (20). 
We wanted to assess whether forcing students to think about
the free-body diagram in each case would help them focus on the similarity of the problems. 
Although the performance improved somewhat when students were also asked about the free body diagrams (Table 1 presents data for
the case when students were given questions (17)-(20)), 
it is not significantly different from when they were only asked question (20). 
The strong misconception prevented transfer of relevant knowledge from the problem
not involving friction to the one involving friction, even when students
were explicitly asked for the free-body diagrams in the two cases.
The most common incorrect response to question (19) was choice (a), because these students believed that the frictional force should
be pointing down the incline. 
Approximately, $40\%$ believed that friction had a magnitude $\mu_{s}N$ and approximately $30\%$ believed it was $\mu_{k}N$.
In individual interviews, students often noted that the problem with
friction must be solved differently from the problem involving tension
because there is a special formula for the frictional force. Even
when the interviewer drew students' attention to the fact that the
other forces (normal force and weight) were the same in both questions
and they are both equilibrium problems, only some of the students
appeared concerned. Others used convoluted reasoning and asserted
that friction has a special formula which should be used whereas tension
does not have a formula, and therefore, a free-body diagram must be used.

In the earlier administration, questions (21) and (22) were given as an IPP but
in the later administration whose results are given in Table 1, either three questions (21)-(23) were given
together as a triplet, or question (22) or question (23) involving friction were given alone.
Table 1 shows that the performance on questions (22) or (23) did not
improve significantly when they were given with question (21).
The most common incorrect response in question (22) was $\mu_{s}N=600 N$
($\sim40\%$), with or without question (21). 

Misconceptions about friction were so strong that students who were given both problems did not fully 
discern their similarity and take advantage of their responses to question (21) to analyze the horizontal 
forces in question (22).
An alternative knowledge-in-pieces view
can also be used to explain these findings in terms of students activating
different resources to deal with somewhat different contexts which
experts view as equivalent. Smith et al.~\cite{smith} argue that
student responses should be considered as {}``resources\char`\"{}
rather than flawed and note that \textit{{}``Persistent misconceptions,
if studied in an evenhanded way, can be seen as novice's efforts to
extend their existing useful conceptions to instructional contexts
in which they turn out to be inadequate. Productive or unproductive
is a more appropriate criterion than right or wrong, and final assessments
of particular conceptions will depend on the contexts in which we
evaluate their usefulness.\char`\"{}} 
From this point of view, the problem context
triggers activation of knowledge that students think is relevant and reach a conclusion
that is incorrect but that nevertheless makes sense to them. Therefore, students
do not feel the need to look further for analogies to the paired problem.
In question (23), the coefficient of friction
was not provided, and similar to the common misconception in question
(20), almost $50\%$ of the students believed that it is impossible
to determine the resultant force on the crate without this information.

Similarly, although the frictional force in question (25) is irrelevant for the question asked, it was
a distracting feature for a majority of students. 
Common incorrect reasoning for question (25) was based on the 
assumption that friction must play a role in determining the work done by the
person and the angle of the ramp was required to calculate this work
even though the distance by which the box was moved along the ramp
was given. Interviews suggest that many students had difficulty distinguishing
between the work done on the box by the person and the total work
done. They asserted that the work done by the person cannot be the
same in the two problems because friction must make it more difficult
for the person to perform the work.

\subsection{Survey about the effectiveness of the IPPs}

In one of the courses in which students were given many of the IPPs,
they were also given a multiple-choice survey with the following questions: 

\begin{enumerate}
\item Did you notice a pairing between the problems on the quiz? (choices:
(a) Yes, it was obvious, (b) yes, after a while, (c) a few questions
seemed paired, (d) maybe one, (e) not at all) 
\item Did the paired problems cause you to reconsider any answers? (choices:
(a) Yes, all of them, (b) several of them, (c) a few, (d) maybe one,
(e) not at all) 
\item Were the paired problems helpful? (choices: (a) The first problem
in a pair helped me with the second, (b) The second problem helped
me with the first, (c) The problems helped me with each other, (d)
They did not help me at all, (e) They were actually confusing) 
\item Which type of problems were most helpful, if any? (choices: (a) Algebraic
answer, (b) Numerical answer, (c) Comparison (more/less), (d) Simple
question, with a reason, (e) Scaling question, e.g., if you double
the radius...) 
\item You had a chance to explain multiple choice answers for partial credit.
Did you find that this helped you formulate the answer better? (choices:
(a) Yes, very much so, (b) helped somewhat, (c) so so, no effect,
(d) didn't help at all, and used up time, (e) it was actually confusing.) 
\end{enumerate}
The survey data is self-reported and should be interpreted with this fact in mind. 
In response to survey question (1), more than $50\%$ of students chose (a), claiming to notice the pairing immediately.
In response to survey question (2), more than $40\%$ of students noted that the
paired problems caused them to reconsider at least a few answers (choice
(c)) and in response to question (3), about $40\%$ of students noted
that the paired problems helped them with each other (choice (c)).
These responses again suggest that students were actually trying to
make sense of the problems to the best of their ability and taking
advantage of the IPPs. In response to question (4), choices (a), (b)
and (d) were all equally popular. These responses are consistent with
the fact that algebraic or numerical problems gave students some confidence
and provided them with tools to make sense of the paired conceptual
problems. In response to question (5), more than $30\%$ of students
noted that explaining the multiple choice answers helped somewhat
in better formulating their responses while $20\%$ noted that it
helped very much.

\vspace*{-0.2in}
 
\section{Summary}

\vspace*{-0.2in}

Student performance on the quantitative problems did not improve significantly when they were 
paired with the corresponding conceptual questions compared to when quantitative problems were given alone.
However, students often performed significantly better 
on the conceptual questions when both quantitative and conceptual questions
were given than when the conceptual question alone was given. 
Individual discussions and written responses suggest that many students were
able to recognize the isomorphisms between problems, reason about their quantitative solution and transfer
that knowledge to the conceptual solution. 

While students often took advantage of the quantitative problem to answer the corresponding
conceptual question of an IPP, those who were only given the corresponding conceptual
question did not automatically convert it into a quantitative problem as an aid for reasoning correctly.
Examination of students' scratch work suggests that they seldom attempted such conversion 
by choosing appropriate variables.
One-on-one discussions suggest that students often used gut feeling to reason about the conceptual questions.
This tendency persisted even when the interviewer explicitly encouraged students to convert a conceptual question into
a quantitative one. It is possible that converting the conceptual questions to quantitative ones was too cognitively 
demanding for introductory students and may have caused mental overload.

Even in IPPs that did not pair quantitative and conceptual questions but one question provided
a hint for the other, students could sometimes exploit the reasoning for one
of the questions to answer the other question when both questions
in a pair were given. The fact that many students could discern
the similarity between the problems and take advantage of their solution
to one problem to answer the other one suggests that their expertise
is evolving. In a survey given to students in one course, they noted
that they often realized the similarity of the paired problems and
sometimes tried to make a connection between the problem pairs to
answer the question that was more difficult in each pair.

In this research, isomorphic problems were given back-to-back, and
the more quantitative question always preceded the conceptual question in an IPP. 
The three IPPs in questions 11-16 were also always given in the same order.
It is possible that the order in which questions were asked and
the proximity of the paired questions in an IPP are major factors in whether students 
will recognize their similarity and transfer relevant knowledge from one problem to another. 
In future research, one can explore the effect of
spacing the isomorphic problems and changing the order, e.g., of the quantitative
and conceptual questions, on students' ability to benefit from having
both questions of an IPP. Changing the order in future research would
also be insightful for the IPPs in which both questions were relatively
conceptual (e.g., the three IPPs involving questions 11-16). This
research may be helpful in understanding whether one question in such
IPPs provides a better hint by explicitly mentioning relevant variables
for answering the corresponding paired question.

From a misconceptions standpoint, 
strong alternative views about friction related to the context
of some of the problems often prevented students from seeing
the underlying similarities between the problems involving friction
and an isomorphic problem that students found easier to solve.
For example, many students believed that the static friction is always at the maximum value, or that the kinetic
friction is responsible for keeping the car at rest on an incline,
or that the presence of friction must affect the work done by you even
if you apply the same force over the same distance. 
In such cases, students appeared to frame the isomorphic problems involving friction and not involving
friction differently and traversed different problem spaces while solving them.
From a knowledge in pieces perspective,
the context when friction is present and prominent triggers activation of knowledge
that students think is relevant (e.g,, the formula for maximum static friction) and
they run with it and reach a conclusion that makes sense to them but that is not correct.
Thus, there is no need to look further to similarities to the paired problem or to
anything else. This latter view is similar to Simon's theory of ``satisficing" where
individuals will only select a few of the large number of possible paths in the problem space 
which are consistent with their expertise in the area, satisfies them, and does not cause a cognitive overload.
If an individual is not an expert in
a domain, it is likely that these paths in the problem space are not the ones that will lead to success.
When students satisfice, there is no need to discern the deep similarity of the paired problems and
transfer their analysis for the problem not involving friction to the one involving friction because
within their world view the solution strategy that comes to their mind after understanding the problem makes sense.

\vspace*{-0.2in}

\section{Instructional Implications}

\vspace*{-0.2in}

Although student performance on quantitative problems did not improve
significantly when such problems were paired with conceptual questions, 
students benefited from quantitative and conceptual problem pairs in answering conceptual
questions. Presenting quantitative and conceptual isomorphic pairs
helped students make conceptual inferences using quantitative tools.
Such problem pairs as part of instruction may help students go beyond
the {}``plug and chug\char`\"{} strategy for the quantitative
problem solving and may give them an opportunity to reflect upon their
solution and develop reasoning and meta-cognitive skills. Solving
these paired problems can force students to reflect upon the problem
solving process and improve their meta-cognitive skills. Helping students
develop meta-cognitive skills can also improve transfer of relevant knowledge from one problem to another.

In cases where the strong alternative views about friction prevented transfer of relevant knowledge, students may
benefit from paired problems only after they are provided the opportunity
to repair their knowledge structure so that there is less room for
these alternative views. 
Instructional strategies embedded within a coherent curriculum that force students to realize that the static 
frictional force does not have to be at its maximum value or that the work done on a box by a person who is applying a 
fixed force over a fixed distance will not depend on friction may be helpful; students may be attempting to include all 
information in a problem statement to answer a question when only some information is relevant.
Asking students to predict what should happen in concrete situations, helping them realize the discrepancy
between their predictions and what actually happens, and then providing
guidance and support to enhance their expertise is one such strategy.~\cite{piaget,vygotsky}.

Isomorphic problems can be exploited as useful tools for teaching
and learning. One strategy is to give isomorphic problems similar
to those in this study and then discuss their isomorphism later with
students to help them learn to discern the underlying similarities
of the problems. Another strategy is to tell students that the problems
are isomorphic and ask them to justify the isomorphism. Using these
strategies with a variety of isomorphic problems with varying difficulty
can help develop expertise and improve students' ability to transfer relevant knowledge
from one context to another. Also, the simplest level of isomorphic
problems where the same problem is asked with different parameters
can be a useful tool for teaching students to do symbolic manipulation.
Unlike the expert strategy, some students may trade the symbols for
numbers in the equations at the beginning while solving problems because
they may not recognize the advantage of symbolic manipulation~\cite{fred}
or may not have the mathematical skills to carry out algebraic manipulations with symbols. 
One hypothesis for future testing is that, if students
are consistently given homework problems where they have to solve
problems with different sets of numerical parameters and they are
told that if they obtain a correct symbolic answer, they will get
full credit without inserting each set of parameters, whether they are motivated to perform symbolic manipulation. 
Another important issue often is one of extracting meaning from symbolic manipulations because
some students can manipulate symbolic equations and yet not be able to interpret the physical 
meaning once the answer is reached. 
Students can be rewarded for identifying isomorphic problems in their homework problems, e.g.,
if they explain why two problems are isomorphic they can only solve
one of them in great detail and can simply lay out the plan for solving the other one.
Such reward policy can motivate students to perform a conceptual analysis and planning before
jumping into the implementation phase of problem solving and can help them extract meaning from mathematical manipulations.

\vspace*{-0.2in}
 
\section{Acknowledgments}

\vspace*{-0.2in}

We are very grateful to J. Mestre for giving extensive feedback on the manuscript.
We thank P. Reilly for help in developing the questions and to him, F. Reif, R. Glaser, R. P. Devaty and J. Levy
for useful discussions and to all faculty who administered the problems
to their classes. We thank the National Science Foundation for award NSF-DUE-0442087.

\pagebreak

\bibliographystyle{aipproc}

\begin{thebibliography}{9}

\bibitem{gick} M. Gick and K. Holyoak, {\it The cognitive basis of knowledge transfer},
in Transfer of learning: Contemporary research and applications, Cornier $\&$ Hagman (Eds.), New York, Academic Press (1987);
M. Gick and K. Holyoak, {\it Schema induction and analogical transfer}, Cognitive Psychology, {\bf 15}, 1-38, 1983;
K. Holyoak, {\it The pragmatics of analogical transfer}, In The Psychology of learning and motivation, Ed. G. Bower,
(Vol. 19), New York: Academic Press, 1985;
K. Holyoak and P. Thagard, {\it Mental Leaps: Analogy in creative thought}, Cambridge, MA: MIT press, 1995.

\bibitem{mestre} R. Dufresne, J. Mestre, T. Thaden-Koch, W. Gerace and W. Leonard, {\it Knowledge representation and coordination
in the transfer process} 155-215, in {\bf Transfer of learning from a modern multidisciplinary perspective}, J. P. Mestre (Ed.), 393 pages,
Greenwich, CT: Information Age Publishing, (2005)

\bibitem{lobato} J. Lobato, {\it How design experiments can inform a rethinking of transfer and vice versa}, Educational Researcher,
32(1), 17-20, (2003); J. Lobato, {\it Alternative perspectives on the transfer of learning: History, issues and challenges for
future research}, J. Learning Sciences, {\bf 15(4)}, 431-439, (2006).

\bibitem{sanjay} D. J. Ozimek, P. V. Engelhardt, A. G. Bennett, N. S. Rebello, {\it Retention and transfer from
trigonometry to physics}, Proceedings of the Physics Education Research Conference, {eds. J. Marx, P. Heron, S. Franklin}, 
173-176, (2004).

\bibitem{bassok} M. Bassok and K. Holyoak, {\it Interdomain transfer between isomorphic topics in algebra and physics}, J. Experimental
Psychology: Memory, Learning and Cognition, {\bf 15(1)}, 153-166, (1989).

\bibitem{bransford} J. D. Bransford and D. Schwartz, {\it Rethinking transfer: A simple proposal with
multiple implications}, Review of Research in Education, {\bf 24}, 61-100, (1999).

\bibitem{sternberg} {\it Transfer on trial: Intelligence, cognition and instruction} 
D. K. Detterman and R. J. Sternberg eds., Norwood NJ: Ablex, (1993);
R. J. Sternberg, {\it Component processes in analogical reasoning}, Psychological Review, {\bf 84}, 353-378, (1977).

\bibitem{transfer5} L. Novick, {\it Analogical transfer, problem similarity and expertise}, J. Experimental Psychology:
Learning, memory and cognition, {\bf 14(3)}, 510-520, (1988). 

\bibitem{analogy_transfer} A. Brown {\it Analogical learning and transfer: What develops?}, in {Similarity and analogical
reasoning}, (S. Vosniadu and A. Ortony Eds.), Cambridge U.P., NY, PP. 369-412, (1989).
E. Kunz and R. Tweney, {\it The practice of math and science: From calculations to the clothesline problems},
In Rational models of cognition, Eds. M. Oalesfard and N. Chater, pp. 415-438, (1998);
P. Adey and M. Shayer, {\it An exploration of long-term far-transfer effects following an extended intervention program
in the high school science curricula}, Cognition and Instruction, {\bf 11}, 1-29, (1993);
D. Genter and C. Toupin, {\it Systematicity and surface similarity in the development of analogy}, Cognitive
Science, {\bf 10}, 277-300, 1986.

\bibitem{hendrickson} G. Hendrickson and W. Schroeder, {\it Transfer training in learning to hit a submerged target},
J. Ed. Psych., {\bf 32}, 205-213, 1941.

\bibitem{klahr} D. Klahr and S. M. Carver, Cognitive objectives in a LOGO debugging curriculum: Instruction, learning
and transfer, Cognitive Psychology, 20, 362-404, 1988.

\bibitem{chi3} M. T. H. Chi, P. J. Feltovich, and R. Glaser, {\it Categorization and
representation of physics knowledge by experts and novices}, Cog. Sci. {\bf 5}, 121-152 (1981).

\bibitem{hardiman} P. T. Hardiman, R. Dufresne and J. P. Mestre, {\it The relation between problem categorization and problem
solving among novices and experts}, Memory and Cognition, {\bf 17}, 627-638, (1989).

\bibitem{lillian} L. C. McDermott, Oersted Medal Lecture 2001: Physics Education Research: The key to student learning,
Am. J. Phys. {\bf 69 (11)}, 1127-1137, (2001);
B. S. Ambrose, P. S. Shaffer and L.C. McDermott 
{\it An investigation of student understanding of single-slit diffraction and double-slit 
interference}, Am. J. Phys. {\bf 67 (2)}, 146-155, (1999).

\bibitem{kim} E. Kim and S-J Pak, {\it Students do not overcome conceptual difficulties after solving 1000 traditional problems},
Am. J. Phys. {\bf 70 (7)}, 759-765, 2002.

\bibitem{eric}  E. Mazur, {\it Peer Intruction: A User's Manual}, Series
title: Prentice Hall series in educational innovation. Upper Saddle River, Prentice Hall, N.J., (1997).

\bibitem{tutorial} C. Singh, {\it Interactive video tutorials for enhancing problem solving, reasoning, and
meta-cognitive skills of introductory physics students}, C. Singh, Proceedings of
Phys. Ed. Res. Conference, Madison, WI, (S. Franklin, K. Cummings, J. Marx Eds.), 2003.

\bibitem{fred2} F. Reif, {\it Millikan Lecture 1994: Understanding and teaching important
scientific thought processes}, Am. J. Phys. {\bf 63}, 17 (1995).

\bibitem{hatano} G. Hatano and K. Inagaki, {\it Two courses in expertise}, in H. Stevenson, J. Azuma and K. Hakuta eds.,
Child development and education in Japan, W. H. Freeman and Co., NY, 262-272, (1986).

\bibitem{rosengrant} Singh and D. Rosengrant, Multiple-choice test of energy and
momentum concepts, Am. J. Phys. 71, 607 (2003). 

\bibitem{disessa} A. A. diSessa and B. Sherin, {\it What changes in conceptual change?},
Int. J. Sci. Ed., {\bf 20(10)}, 1155-1191, (1998).

\bibitem{fred} F. Reif, {\it Teaching problem solving-A scientific approach}, The Phys. Teach. {\bf 33}, 310, (1981);
J. H. Larkin, and F. Reif, {\it Understanding and teaching problem solving in physics},
Eur. J. Sci. Ed. {\bf 1(2)}, 191-203, (1979).

\bibitem{sweller} J. Sweller, {\it Cognitive load during problem solving: Effects on learning},
Cognitive science {\bf 12}, 257, (1988); J. Sweller, R. Mawer, and M. Ward, {\it Development
of expertise in mathematical problem solving}, J. Exptal. Psychology: General {\bf 112}, 639, (1983).

\bibitem{simon} 
H. A. Simon, {\it Reason in human affairs}, Basil Blackwell, (1983);
A. Tversky, and D. Kahneman, {\it Judgement under uncertainty: Heuristics and biases}, Science, {\bf 185}, 1124-1131, (1974);
J. Reason, {\it Human Error}, Cambridge Univ. Press, (1990).

\bibitem{friction} C. Singh, {\it Effect of Misconceptions on Transfer in Problem Solving}, 
AIP Conf. Proc. {\bf 951}, 196-199, (2007).

\bibitem{smith} J. Smith, A. diSessa and J. Rochelle, {\it Misconceptions reconceived-A constructivist analysis of knowledge
in transition}, J. Learning Sciences, {\bf 3(2)}, 115-163, (1993).

\bibitem{piaget} H. Ginsberg, and S. Opper, {\it Piaget's theory of intellectual development},
Englewood cliffs, Prentice Hall, N.J., (1969); R. Gorman, {\it Discovering Piaget:
A guide for teachers}, Merrill, Columbus, (1972).

\bibitem{vygotsky} L. Vygotsky, {\it Mind in Society: The development of higher psychological processes}, Harvard University Press, 1978; J.
Wertsch, {\it Mind in Action}, NY: Oxford University Press, (1998).

\end{thebibliography}

\pagebreak

\begin{table}[h]
\centering
\begin{tabular}[t]{|c|c|c|c|c|}
\hline
Problem $\#$ & only one & both & p value&Chi square\\[0.5 ex]
\hline \hline
1& 59 (138)&54 (289)& 0.40 &0.8 \\[0.5 ex]
2& 31 (215) & 58 (289)&0.00& 36.0\\[0.5 ex]
\hline
3& 34 (138) &38 (289)& 0.45 & 0.6 \\[0.5 ex]
4& 23 (215) &30 (289) &0.07 & 3.3 \\[0.5 ex]
\hline
5& 81 (138) &76 (289) &0.26&1.4\\[0.5 ex]
6& 55 (215) &80 (289) &0.00&36.3\\[0.5 ex]
\hline
7& 52 (138) &56 (289)& 0.47 & 0.6\\[0.5 ex]
8& 44 (150) & 51 (289)& 0.19& 1.9 \\[0.5 ex]
\hline
9 & 49 (138)& 49 (289) & 1.00& 0.0 \\[0.5 ex]
10 & 53 (150)& 71 (289) & 0.00 &13.4 \\[0.5 ex]
\hline
11 & 53 (81) &65 (289) &0.05 &3.9\\[0.5 ex]
12 & 23 (65)& 52 (289) & 0.00 & 17.7 \\[0.5 ex]
\hline
13 &50 (81) & 52 (289) &0.9 &0.0\\[0.5 ex]
14 & 33 (65)&58 (289)& 0.00& 14.2 \\[0.5 ex]
\hline
15 & 65 (81) & 74 (289) & 0.16 & 2.3\\[0.5 ex]
16 & 64 (65) & 86 (289) & 0.0 & 16.1 \\[0.5 ex]
\hline
17 & & 90 (479) & &\\[0.5 ex]
19& & 67 (479) & &\\[0.5 ex]
\hline
18  & & 72 (479) & &\\[0.5 ex]
20 & 20 (81)& 28 (479) & 0.14 & 2.4 \\[0.5 ex]
\hline
21  & & 77 (150)& & \\[0.5 ex]
22  & 24 (190) & 30 (150) & 0.27 & 1.4 \\[0.5 ex]
23  &18 (81)  & 16 (150)& 0.71 &0.2  \\[0.5 ex]
\hline
24  & & 71 (150)  & & \\[0.5 ex]
25  & 30 (81) & 32 (150) & 0.77& 0.1\\[0.5 ex]
\hline
\hline
\end{tabular}
\vspace{0.1in}
\caption{For the isomorphic problems given in the multiple-choice format (see the Appendix), the first column lists
the problem numbers, the second column gives the percentage of students who chose the correct answer when
only one of the questions was given to them, and the third column gives the percentage of students who chose
the correct answer when both questions (triplet for questions (21)-(23)) were given. 
The numbers in parentheses in the second and third columns refer to the number
of students who answered the question. The last two columns for all questions 
list the p value and Chi square for comparison of student performance between cases when only one of the 
isormorphic questions was given vs. when the question was given with its isomorphic pair. In experiment 3,
we only test for significant differences for questions involving friction when they were given alone vs. with an
isomorphic question not involving friction. Questions (18) and (20) are isomorphic but they are not consecutive because 
for the results presented in the table,
they were given with the corresponding free body diagrams (i.e., students who answered both questions (18) and (20),
actually answered questions (17)-(20) in that order).
}
\label{junk}
\end{table}

\end{document}